\documentclass[reprint,amsmath,amssymb,aps,twocolumn,superscriptaddress,nofootinbib]{revtex4-2}

\pdfoutput=1

\usepackage{footnote}
\usepackage{graphicx}% Include figure files
\usepackage{mathtools}
\usepackage{dcolumn}% Align table columns on decimal point
\usepackage{bm}% bold math
\usepackage{float}
\usepackage[utf8]{inputenc}
\usepackage{booktabs}
\usepackage{tabularx}
\usepackage{colortbl}
\usepackage{etoolbox}
\usepackage{titlesec}
\usepackage{braket}
\usepackage{xcolor}
\usepackage[font=small]{caption}
\usepackage{footnotebackref}
\usepackage{tablefootnote}

\titlespacing*{\section}{0pt}{0.5\baselineskip}{0.4\baselineskip}
\titlespacing*{\subsection}{0pt}{0.5\baselineskip}{0.4\baselineskip}

\begin{document}

\preprint{APS/123-QED}

\title{Can Traditional Terrestrial Applications of Gravity Gradiometry Rely Upon Quantum Technologies? A Side View}

\author{Alexey V. Veryaskin}
\email{alexey.veryaskin@uwa.edu.au}
\affiliation{Quantum Technologies and Dark Matter Laboratory (QDM Lab, UWA), Department of Physics, University of Western Australia, 35 Stirling Highway, Crawley, WA 6009, Australia.}
\affiliation{Trinity Research Labs, Department of Physics, University of Western Australia, 35 Stirling Highway, Crawley, WA 6009, Australia.}
\author{Michael E. Tobar}
\email{michael.tobar@uwa.edu.au}
\affiliation{ARC Centre of Excellence for Engineered Quantum Systems and ARC Centre of Excellence for Dark Matter Particle Physics, Department of Physics, University of Western Australia, 35 Stirling Highway, Crawley, WA 6009, Australia.}
\affiliation{Quantum Technologies and Dark Matter Laboratory (QDM Lab, UWA), Department of Physics, University of Western Australia, 35 Stirling Highway, Crawley, WA 6009, Australia.}
\date{\today}

\begin{abstract}

The era of practical terrestrial applications of gravity gradiometry begun in 1890 when Baron Lorand von Eötvös, a Hungarian nobleman and a talented physicist and engineer, invented his famous torsion balance – the first practical gravity gradients measuring device. It was credited for the major oil discoveries later in Texas (USA). A 100 years later Kasevich and Chu pioneered the use of quantum physics for gravity gradient measurements. Since then cold-atom gravity gradiometers, or matter-wave gravity gradiometers, have been under development at almost every physics department of top-rated universities around the globe. After another 30 years since the Kasevich and Chu publication of 1992, which had led to the first ever quantum gravity gradiometer, the corresponding research and development ceased from being profoundly active a few years back. This article is an attempt to understand and explain what may have happened to the Quantum invasion into the area of applied physics and precision engineering that traditionally has been occupied by non-quantum technologies developed for about a 130 years of the history of gravity gradiometry.

\end{abstract}


\maketitle

\section{Introduction}

Gravity gradients constitute the second spatial derivatives of the gravitational potential and provide valuable information about the presence and/or nature of underground/subsea oil, gas and mineral resources, and assist with navigation, defence and several other areas of industrial, commercial and technological endeavour \cite{Veryaskin2018}. There is a significant number of different approaches to measuring gravity gradients, which represent a second rank tensor having five independent components (out of nine) that can be measured individually or concurrently. Among the recent ones are a resurrection of the 100+ years old torsion balance \cite{Zhang:2020wt}, micromachined (MEMS) gravity gradiometers \cite{Marocco19} and quantum or matter-wave gravity gradiometers \cite{Hu:2019uf}.

The first quantum gravity gradiometer, based on laser-manipulated atom interferometry, was developed at Mark Kasevich’s laboratory, based at Yale University \cite{Snadden98}. The device used a light-pulsed magneto-optical atom trapping technique to simultaneously measure the absolute accelerations of two proof masses, each free falling along a ballistic trajectory. Each proof mass was a laser-cooled ensemble (cloud) of Ceasium (Cs) atoms. The gravity gradient was then extracted by dividing the difference of the two acceleration measurements by the baseline distance between the two ensembles (set between 1 and 10 m).

Since the cornerstone article of Kasevich and Chu on using light-matter interaction was published about 30 years ago \cite{Kasevich:1992wn}, there have been a tremendous rise in the number of on-line presentations and academic publications, including leading science journals, promoting so-called quantum technologies and quantum sensing without explaining much what exactly they mean and sometimes without much experimental support to the claimed benefits. Quantum gravity gradiometers, or matter-wave gravity gradiometers in other words, had been under development at almost every physics department of top-rated universities around the globe. However, in the last couple of years the number of publications containing the words “cold-atom gravity gradiometers” has reduced to about zero\footnote{The authors believe the COVID-19 Pandemic is not the main factor that caused this to have happened.}. Also, gravity gradiometers which use the matter-wave interferometry as the founding principle, have vanished from the list of potential strategic terrestrial applications and moved to such fundamental physics areas as gravitational wave detection \cite{Canuel:2020uc}, the measurement of the gravitational constant (the big-G) \cite{Rosi:2014tp,Mao:2021ue} and some others\cite{Parker:2018tv}. Long base-line matter wave interferometry and quantum gravity gradiometers designed for future space missions is still an active research \cite{Muller:2020vy,Abe_2021}, where matter-wave interferometry applied to gravity gradient measurements is still under consideration \cite{Muller:2020vy}. A recent review \cite{Tino_2021} provides a detailed picture of the research areas where cold-atom interferometry has been used for fundamental physics experiments and for measuring gravitational acceleration and gravity gradients. 

In contrast, the latest publication that considered out-of-the-lab applications was in 2019 \cite{Bongs:2019uj}.  Bongs and co-authors \cite{Bongs:2019uj} described the potential applications of interferometric quantum sensors where only one application of a quantum gravity gradiometer, designed and tested at the University of Birmingham (UK), was mentioned, namely – civil engineering. There were many more on the list but rather traditional areas where non-quantum gravity gradiometers have been used successfully. Civil engineering is not very much a traditional niche for gravity gradiometry. It is well known from the history of gravity gradiometry, the first gravity gradiometers were credited for oil discoveries in the 1930s of the last century and for the aid to navigate nuclear strategic submarines later in the 1960s. The passive non-jammable navigation, based on mapping fine signatures of the earth’s gravity, airborne mineral exploration and borehole gravity gradiometry are still the most demanding traditional terrestrial applications of gravity gradiometers.

Since the article \cite{Bongs:2019uj} was published in Nature Review (Physics) in 2019, there have been no reports on any advances in using matter-wave based gravity gradiometers for traditional commercial and defence related applications. There also have been some misleading considerations made recently in regard of using quantum gravity gradiometers. In a recently published report titled “Quantum Technology and Submarine Near-Invulnerability” \cite{Kubiak} the author considers using them for detecting submarines. The author is right saying that the mass of a submarine, as any other mass, creates a gravitational signature which, in theory, can be measured by a very sensitive quantum gravity gradiometer without explaining what the very sensitivity should be provided. However, no matter what the mass distribution is in the submarine’s hull, the averaged density of floating subs is equal to the density of sea water due to the elementary buoyancy effect. This means there will not be any detectable gravitational anomaly seen at distances more than a few submarine’s lengths as the latter can only be created by density contrast. The report is an important discussion about oceans’ transparency and the author’s final conclusion that quantum gravity gradiometers will not make oceans fully transparent or seriously endanger submarine near-invulnerability is correct.

There have been a number of commercial spin-offs created since the Kasevich and Chu first publication in 1992 which has led to the first ever quantum gravity gradiometer. Two of them - AOSense Inc. (USA) and Muquans (France) have been among the first commercial start-ups offering quantum sensing for sale.. There are a number of high-tech products, including quantum gravimeters and inertial sensors, available for sale there but no gravity gradiometers based on cold atom interferometry. Muquans has sales offices in China and India, which means the gravimeters does not fall into the ‘dual-use category’ of products that require export permission. 

Singapore-based Atomionics offers the same kind of products and another cold-atom based gravimeter nicknamed GravioTM. In a recent publication \cite{Kumar:2020ug} the authors discuss the advantages of using such quantum technologies as cold-atom and matter-wave interferometry and their possible applications in a similar fashion as in \cite{Bongs:2019uj}.

In Australia, an Australia National University (ANU) based start-up, Nomad Atomics, “seeing the limitations of current commercially available quantum sensors – is developing its own compact, low cost cold-atom gravimeters to combat the problem”. No quantum gravity gradiometers are mentioned on their website despite their parental ANU’s Department of Quantum Science had aimed at developing cold-atom gravity gradiometers a few years back.

There are some other companies around the globe specialising in quantum technologies \cite{Dargan2021} but none of them offer quantum gravity gradiometers for sale after about 30 years from the first attempt to create such instruments. Below is an attempt to understand and explain what happens to the quantum invasion in the area of applied physics and precision engineering that traditionally has been occupied by non-quantum technologies developed for about 130 years of the history of gravity gradiometry.

\section{Quantum Test Masses and Quantum Sensing}

The only physical principle that describes the interaction between the field of gravity and matter is Newton’s law, as the terrestrial approximation of the Einstein’s General Relativity. On earth, a single atom and a brick will fall to ground in exactly the same way. They both represent what is called a test mass concept widely used in accelerometers, gravimeters and gravity gradiometers. The field of gravity forces matter to move. If there are no constraints or/and boundary conditions that restrict an object to fall free, then the motion will continue in free fall. Since the motion is mechanical another mechanical element may be added to balance the motion due to gravity. A typical example is a test mass attached to a spring – this is the core principles of making accelerometers and gravimeters. They represent so-called relative instruments – if the force of gravity changes the test mass will change its equilibrium position due to the restoring force of the spring. The latter is directly proportional to a spatial difference in the strength of gravity at different earth’s locations or a temporal change at a single point (earth tides, as an example).

If one uses a single atom as a test mass and replaces the mechanical spring by a magneto-optical trap (MOT) then this will represent an atom-based gravimeter (not absolute) or accelerometer, but the principle is exactly the same. The only difference is that the test mass is now a quantum-mechanical object. In order to create quantum absolute gravimeter, one need to use atoms (practically clouds of atoms) as freely falling quantum objects. However, in order to use in full their quantum nature, one needs to cool them down to almost absolute zero temperature, keep in a high vacuum and use extremely stable lasers in order to realise sensitive enough matter-wave interferometer – the complications that would significantly increase the cost of making a portable and small enough practical instruments. In both cases of either classic non-quantum test masses or quantum test masses one needs to use their relative or free fall motion under the force of gravity in order to get a measure of the latter. In both cases this is done by using some kind of sensing that translates the change in a test mass position, energy state and/or phase difference of wave-functions to a measure of the force of gravity around. The test masses are often called as primary sensing elements. We have four different cases here: \\ \\
A. Classic test masses and classic sensing (capacitive, inductive, optical); \\ \\
B. Classic test masses and quantum sensing by using the sensing elements that work on quantum-mechanical principles. Typical examples are Superconducting Quantum Interference Devices – SQUIDs, that have been widely used in superconducting gravimeters and gravity gradiometers for decades; \\ \\
C. Quantum test masses and classic sensing (unknown at present time); \\ \\
D. Quantum test masses and quantum sensing. \\ \\
The latter case is what one can call as the Quantum Invasion. 

In recent years, the latter case is frequently marked as the invasion of quantum technologies into the practical applications where the latter had never been used before. Let’s elaborate a bit more on that. Firstly, let’s try to understand what the word “quantum” means. In the most general terms it means that something demonstrates a kind of behaviour that has no analogy in the classic world. In Oxford English Dictionary it is described as “a discrete quantity of energy proportional in magnitude to the frequency of the radiation it represents”. Also, quantum means quantum duality, i.e. quantum objects are described by their wave functions and behave like waves. Quantum-mechanical description of the latter gives only likelihoods of such classic characteristics as position, momentum and energy states. In contrary, classic objects are fully characterised by their centre of mass’ position, speed and acceleration.

Quantum may mean single quantum particles like electrons, neutrons, protons, high energy photons. It may mean quantum fluids like super helium and Cooper pairs in superconductors. It may mean multiple features like quantised energy or momentum states in solids. A single transistor or an operational amplifier are quantum devices as their operational principles are based on the tunnel effect – a pure quantum-mechanical effect when electrons move through not conducting barriers due to the interference of their quantum-mechanical wave functions. So, quantum technologies have been with us for a while already – in our smart phones, in TV sets, in computers (but not quantum computers – which are a genuine representation of the Quantum Invasion). 

\begin{figure}[t]
\includegraphics[width=1.0\columnwidth]{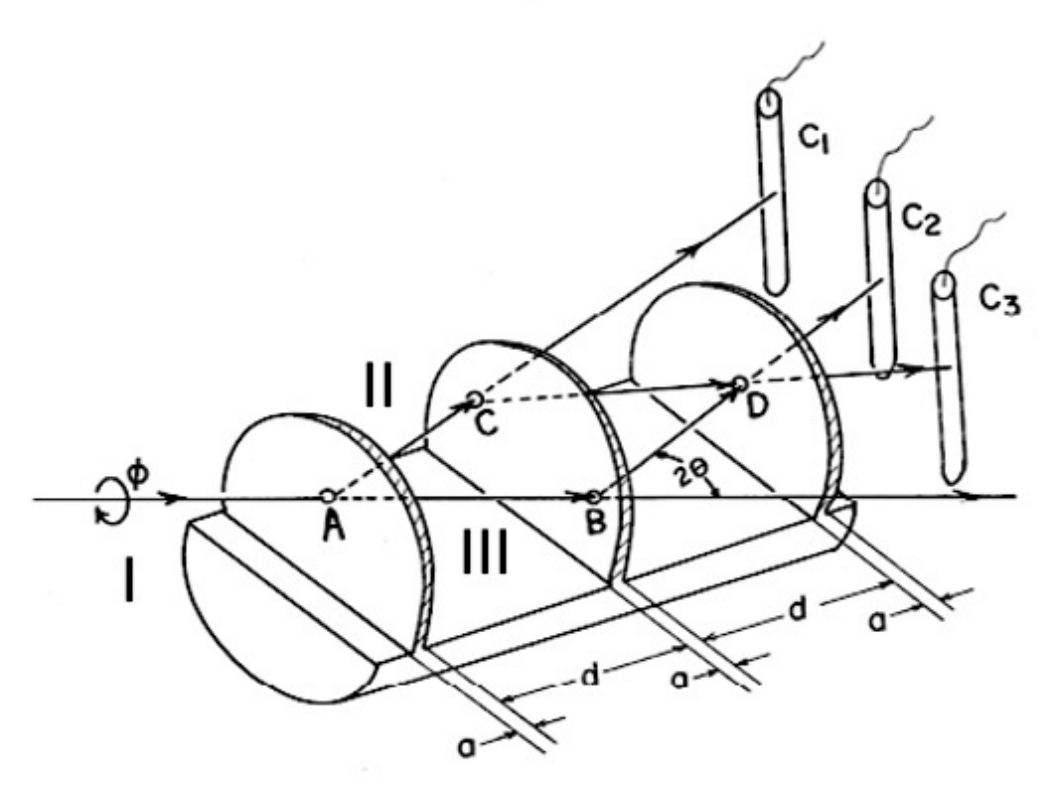}
\caption{Schematic diagram of the neutron interferometer and 3He detectors used in this experiment. The interferometer plane was rotating around the horizontal plane causing the neutron’s quantum-phase interference to be modulated by the projection of the local gravitational acceleration. Reproduced from Colella et al \cite{Colella1975} with permission from American Physical Society.}
\label{NeutInt}
\end{figure}

\section{Matter-Wave Interferometry}

A distinct feature of something that is described by a wave function is that any wave has two parameters, namely, amplitude and phase. The phase can be measured in matter-wave interferometers and be dependent on the strength of the field of gravity around \cite{Overstreet2021}. Effectively only the difference between phases of two matter-waves is measured as is normally done in all interferometric measurements. The first such measurement using a matter-wave interferometer was done by Colella, Overhouser and Werner in 1975 \cite{Colella1975}. They observed a quantum-phase interference of coherent neutron beams propagating in the Earth’s gravitational field and demonstrated that a gravitational potential coherently changes the phase of a neutron’s wave function. The scheme of the experiment is shown in Fig.\ref{NeutInt}. A neutron beam I is split into two coherent beams II and III.

The beam II moves along a trajectory ACD while the beam III moves along a trajectory ABD with a potential energy shift with respect to the beam III. This potential energy difference creates the proportional phase shift between the beams’ wave functions. It is important to underline that there must be a spatial distance between the two neutron beams’ trajectories ACD and ABD where the gravitational potential is different along the travel path. In other words, any interferometric device that measures phase difference generated by the gravitational potential is a phase gradiometer and the result of such measurements is a gradient of the gravitational potential, i.e. a gravitational acceleration vector component.

This has a dramatic impact on the gravity gradiometers which use the matter-wave interferometry. One must establish firstly a distance (base-line) between the arms of such interferometer (typically a Mach-Zehnder one), no matter what they are made of, that can result in a measurable phase gradient and get the local gravitational acceleration (the small-g). Secondly, to use doubled interferometers split by a secondary base-line, in order to get a gravity gradient. In gradiometric measurements it is called a third order gradiometer compared to the first order one that measures directly a desired gravity gradient. The sensitivity of such higher order gradiometers falls much steeper with the distance from mass density contrasts which are the source of gravity gradients. Fig.\ref{Comp} below shows a comparison between two combined atomic interferometers as a gravity gradiometer with a base line $\Delta Z$, and a superconducting third-order magnetic gradiometer comprising two first order gradiometers having a base-line, $d$. Classic test masses do not require any spatial characteristic, such as base-line, in order to measure the gravitational acceleration for the first instance and only require the latter for gradiometric measurements when two test masses are separated by a base-line. Quantum cold-atom or matter-wave based sensors require such a parameter that determines their sensitivity to the primarily sensed quantity, namely the gravitational acceleration. It is also worth noting that if gravitational acceleration is measured as the difference of gravitational potential at separated spatial locations, it is no longer a local quantity.
\begin{figure}[t]
\includegraphics[width=1.0\columnwidth]{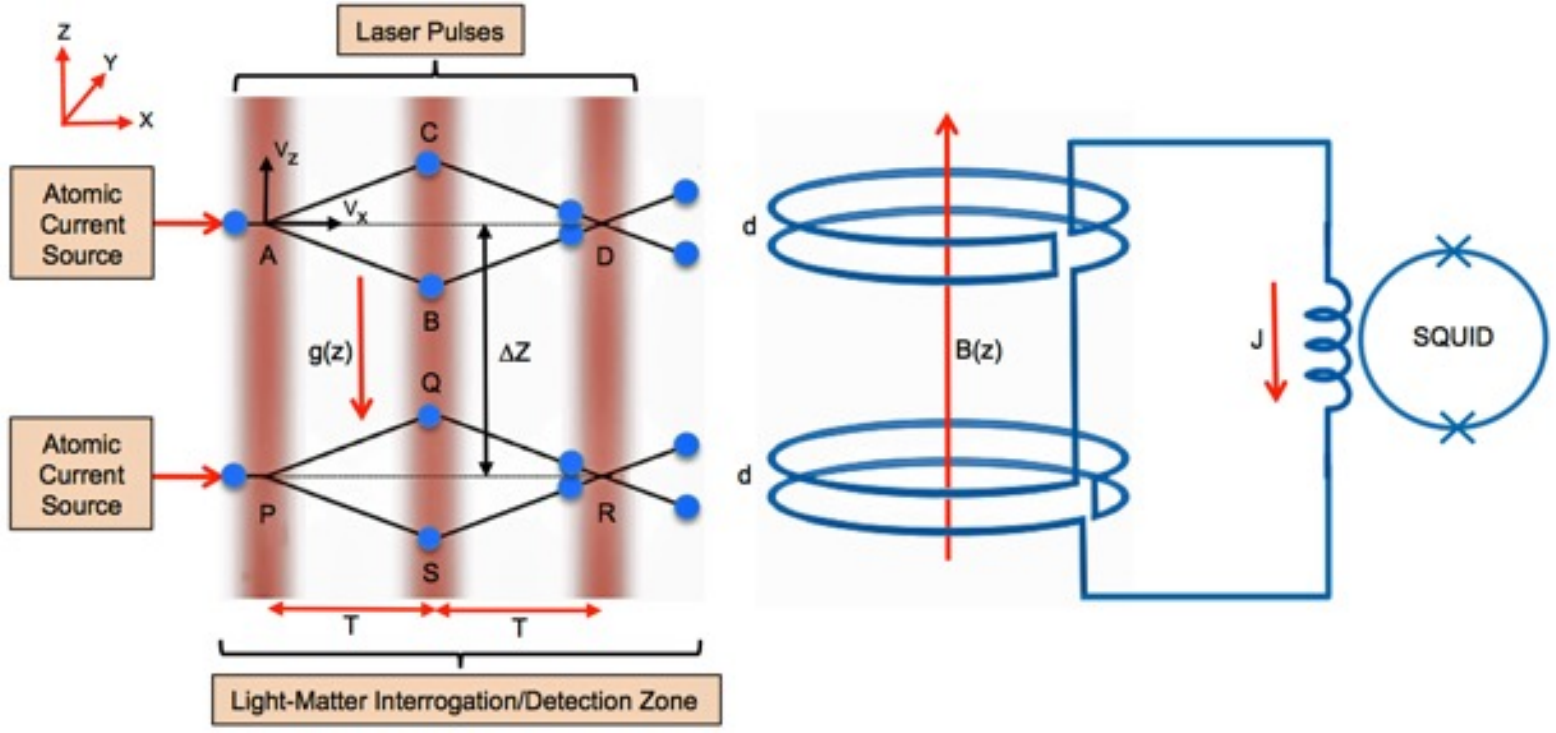}
\caption{A comparison between two combined atomic interferometers as a gravity gradiometer with a base line $\Delta Z$, and a superconducting third-order magnetic gradiometer \cite{Donaldson1984} measuring the third vertical derivative ($B_{zzz}d\Delta Z$) of the magnetic field $B(z)$ and comprising two first order gradiometers having a base-line, $d$.}
\label{Comp}
\end{figure}

Let us consider a Toy-Model of an atomic gravity gradiometer (shown on the left side of Fig.\ref{Comp}) where the real laser pulses, that drive the atomic clouds to their final destinations, are replaced with “magic” laser pulses without getting into the fine details of their low-level interaction with matter. \footnote{For those interested in such low-level quantum-mechanical descriptions of the matter-wave interferometry, see \cite{Storey1994,SANZ2015,Overstreet2021}.}

An atomic cloud, cooled down to about a few micro-Kelvin, is launched from an atomic current source and split at point A by the first laser pulse into two coherent test masses, $M$. The latter travel along the interferometer arms $ACD$ and $ABD$, laid in the $XOZ$ plane, with initial velocities $V_z$ and $V_x$ before being recombined at points ($D,R$) and the phase difference of their wave-packets is extracted by the third laser pulse. The intermediate laser pulse is used to redirect the atomic clouds to their recombination and keep the velocities the same as at the launch point. The time intervals between the laser pulses are the same and equal to $T$.

Let U be the gravitational potential energy, then its contour (clockwise) integral over the interferometer’s area is as follows,
\begin{equation}
\oint_{A C D B A} d \vec{r} U(x, z)=\int d \vec{s} \times \vec{\nabla} U=-M g a \vec{n}_{x},
\label{ACDBA1}
\end{equation}
where g is the absolute value of the local (upper interferometer) gravitational acceleration, $\vec{n}_x$ is a unity vector along the $X$ axis and a is the interferometer’s area that is assumed to be a parallelogram $ACDB$
\begin{equation}
a=\frac{1}{2} B C \times A D=\frac{1}{2}\left(2\left|V_{z}\right| T\right) \times\left(2 T V_{x}\right)=2 V_{x}\left|V_{z}\right| T^{2}
\label{ACDBA2}
\end{equation}
As the vector integral in the left side of Eq.1 has only the X component, it can also be written as,
\begin{equation}
\oint_{A C D B A} d x U(x, z)=V_{x} \oint_{A C D B A} d t U(t)=-\hbar V_{x} \delta \varphi,
\label{ACDBA3}
\end{equation}
where $\delta \varphi$ is the phase shift between the upper and lower interferometer’s arms \cite{Storey1994}
\begin{equation}
\delta \varphi=-\frac{1}{\hbar} \oint_{A C D B A} d t U(t)
\label{dphi}
\end{equation}
One finds from Eq.\ref{ACDBA1}, Eq.\ref{ACDBA2} and Eq.\ref{ACDBA3},
\begin{equation}
\begin{aligned}
&-M g a=-\hbar V_{x} \delta \varphi \\
&\delta \varphi=g \frac{2 M\left|V_{z}\right|}{\hbar} T^{2}=g k_{e f f} T^{2}.
\end{aligned}
\label{dphi2}
\end{equation}
The gravity gradient can be derived from the difference in the phase shifts in the upper and lower interferometers, assuming they are identical,
\begin{equation}
\delta \varphi_{\text {upper }}-\delta \varphi_{\text {lower }}=\Gamma_{z z} k_{e f f} T^{2} \Delta Z
\label{dphi3}
\end{equation}
where $\Gamma_{z z}$ is the vertical gravity gradient component, $k_{e f f}$ is the interferometer’s effective wave vector and $\Delta Z$ is the secondary base line of the gravity gradiometer. The product $k_{e f f} T^{2} \Delta Z$ has the dimension of inverse gravity gradient and one can express the minimum detectable gravity gradient as,
\begin{equation}
\delta \Gamma_{z z}=\Gamma_{0} \sqrt{\left\langle\left(\delta \varphi_{\text {upper }}-\delta \varphi_{\text {lower }}\right)^{2}\right\rangle}, \quad \Gamma_{0}=\frac{1}{k_{e f f} T^{2} \Delta Z}
\label{Gamma}
\end{equation}
where $\Gamma_{0}$ can be treated as the intrinsic gradiometer’s gravity gradient scale factor. The less its value is the more sensitive gradiometer is.
The result in Eq.\ref{Gamma} is the well-known one obtained from the low-level quantum mechanical analysis of real atomic interferometers where the atomic clouds follow ballistic trajectories,
\begin{equation}
z(t)=z_{0}+V_{z} t-\frac{1}{2} g t^{2}.
\label{Ballis}
\end{equation}
The effective area of any such matter wave interferometers (the area within the two split matter waves before they meet again at the point of interference) cannot be made large without breaking the necessary coherence of the two split quantum-mechanical objects and typically limited by tens of millimetres in cold-atom based gradiometers. This results in unavoidable limitation on making compact practical gradiometers if one wants to get a valuable sensitivity level. A doubled-loop single atom interferometer has been described in \cite{Perrin2019}, which allows for a direct gravity-gradient dependent phase shift measurement insensitive to dc acceleration and constant rotation rate. As one would expect, the obtained result had demonstrated a significantly reduced sensitivity of such gradiometer due to the interferometer’s short base-line and a short-term sensitivity of $65,000~E/\sqrt{Hz}$ was obtained during two days of measurements ($1 E = 10^{-9} 1/s^2$ is the unit of gravity gradients  ).

\begin{figure}[t]
\includegraphics[width=1.0\columnwidth]{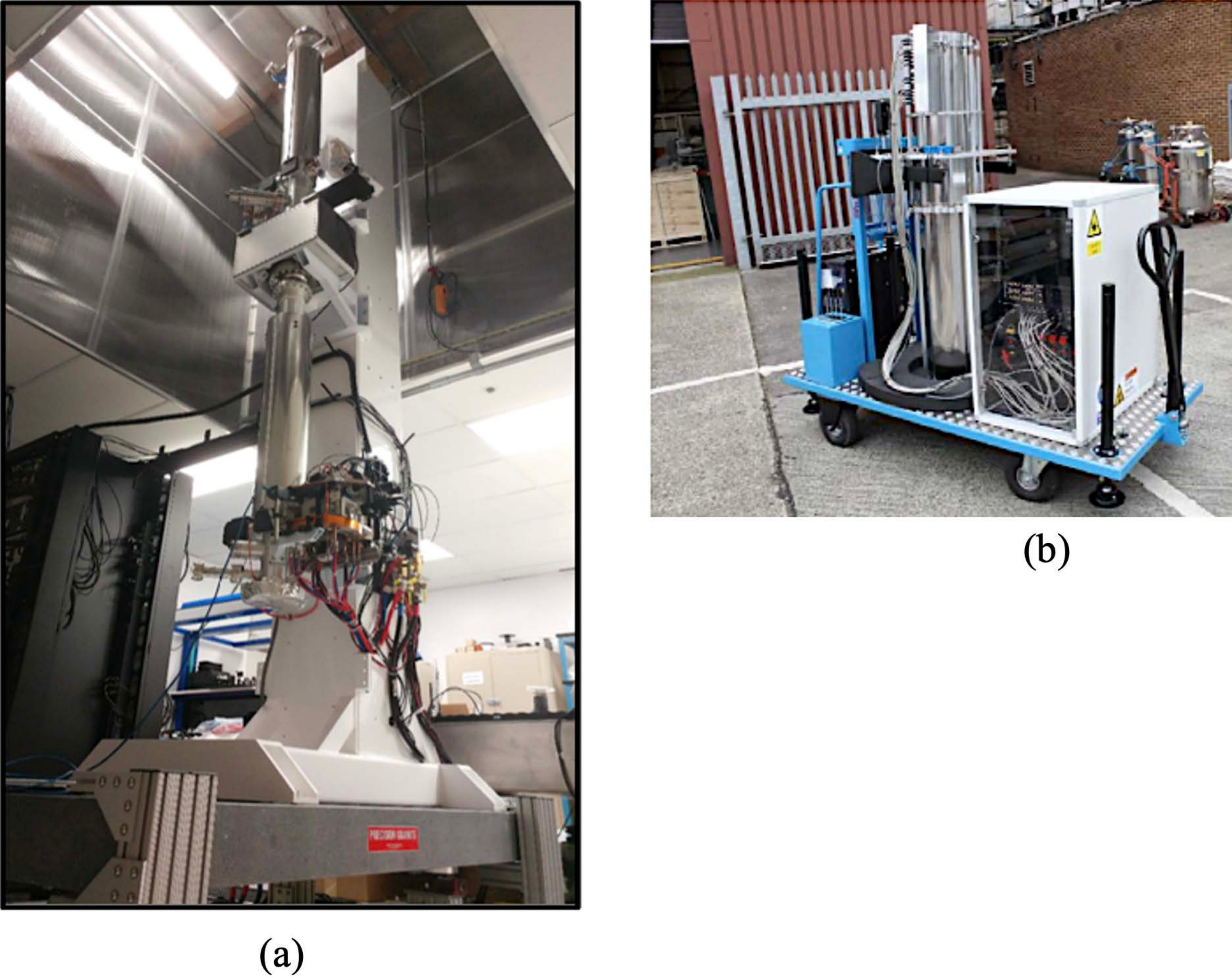}
\caption{(a) a cold-atom gravity gradiometer developed by the AOSense Inc. for space missions, Credit: NASA Earth Sciences Division; (b) a cold-atom gravity gradiometer developed at the University of Birmingham (UK) for civil engineering applications, Credit: Teledyne e2v - REVEAL Ground Based Gradiometer (in collaboration with University of Birmingham)}
\label{Birm}
\end{figure}

The most recent implementation of a cold atom gravity gradiometer that demonstrates the sensitivity of $\sim 100 E/\sqrt{Hz}$, with $0.3$ metres interferometers’ separation base-line has been reported by a Chinese team of scientists from the Key Laboratory of Fundamental Physical Quantities Measurement at the Huazhong University of Science and Technology \cite{Mao:2021ue}. \footnote{This particular gradiometer has been designed for the laboratory use (a big-G measurement)} This result is comparable with the first best results ( $\sim 30-40 E/\sqrt{Hz}$ at $\sim 1$ metre interferometers’ separation base-line) reported by Kasevich’s Group in \cite{Biedermann2015} and in \cite{Sorrentino2014}. Of course, the secondary base-line (the distance between two interferometers) can be made as long as one would have enough space for the gradiometer set up to fit in. Two examples, are shown in Fig.\ref{Birm}(a,b) below. One (a) is a prototype cold-atom gravity gradiometer developed by the AOSense Inc. and funded by NASA \cite{Sugarbaker2018}. The image shows the AIGG instrument in the laboratory with its two metre baseline and two gravimeter sensor heads. The large white structure is not part of the instrument, but a mechanical structure to hold the instrument in the laboratory. While the instrument has a two metre baseline its other dimensions are relatively modest enabling the instrument to be housed in a common spacecraft bus. Another one (b) is a cold-atom gravity gradiometer developed at the University of  Birmingham (UK) for civil engineering applications.

A flight capable cold atom gravity gradiometer is described in \cite{Weiner2020}. This development was aimed at mounting two cold atom interferometers, acting as atomic gravimeters and separated by 1 metre vertical base-line, upon a 6-rotor unmanned arial vehicle (UAV) having a flight time of about 30 min. The total gradiometer payload was expected not to exceed 70 kg. The small-g sensitivity of each interferometer was reported to be $37\times10^{-9} g/\sqrt{Hz}$. The latter was achieved during mobile gravity surveys in the Berkeley Hills (California, USA) along a route of ~7.6 km and an elevation change of ~400 m \cite{Xuejian:uv}.

\begin{figure}[t]
\includegraphics[width=1.0\columnwidth]{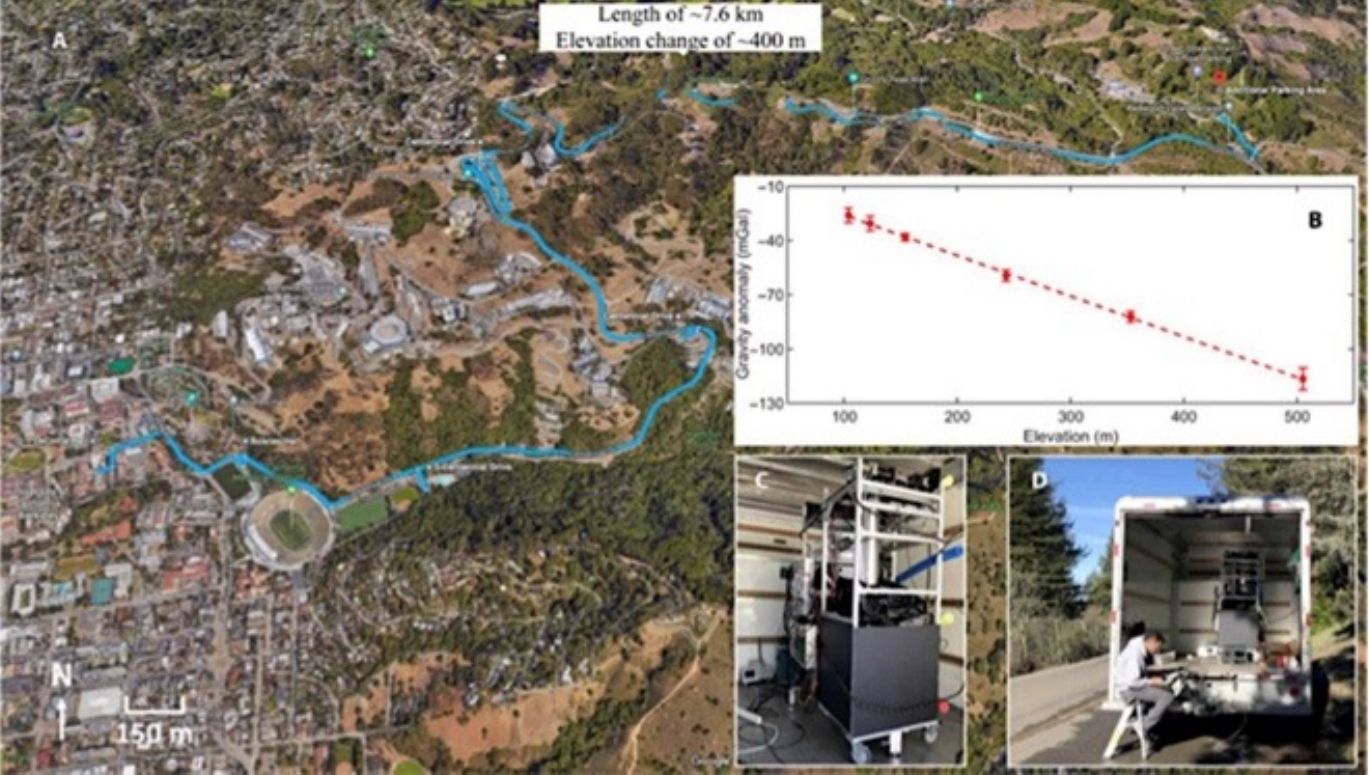}
\caption{Gravity survey in Berkeley Hills. (a) measurement route, the blue curve depicts the route, and the white pin drops are the six measurement locations; (b) gravity anomaly as a function of the elevation; (c) the atomic gravimeter apparatus; (d) field operation of the atomic gravimeter inside a vehicle. Reproduced with permission from Xuejian Wu and the AAAS \cite{Xuejian:uv}. Photo credit for (a): Google Maps; photo credit for (c) and (d): Xuejian Wu, UC Berkeley}
\label{Berk}
\end{figure}

The interferometer was set up in a track as shown in Fig.\ref{Berk}. The vehicle stopped while taking a measurement at different locations and this took about 15 min to set up the gravimeter and a few minutes to measure gravitational acceleration with an uncertainty of around $4\times10^{-8} g$. This translates into the estimate of the doubled interferometer (i.e. gradiometer) noise limit of $360 E/\sqrt{Hz}$. The results of the proposed UAV-based flight test have never been reported in the public domain.

In a recent publication \cite{Stray:2021by} a research team from the University of Birmingham discussed an application of gravity gradiometry to gravitational cartography. They claimed a 20 Eotvos resolution of a cold atom quantum gravity gradiometer over 10 min averaging time and a vertical baseline of 1 m. The instrument's short-term noise level was reported to be about $470~E/\sqrt{Hz}$. There has been also suggested that the technology has the potential to provide a further 10 to 100 fold improvement in instrument sensitivity, allowing faster mapping or detection of smaller and deeper features. It is expected that such performance will be achieved in practical instruments within the next 5-10 years.

\section{Cold Atoms Versus Bose-Einstein Condensate}

In recent years, there have been obvious departures from about 30 years old traditional cold-atom interferometry (CAI) applied to gravity gradiometry \cite{Travagnin2020}. Bose-Einstein Condensate (BEC) based gravity gradiometers may come forward in the nearest future and take over the traditional CAI instruments \cite{Szigeti2021}. BEC based instruments can still be realised as matter-wave interferometers. \footnote{A phononic gravity gradiometer using collective oscillations of trapped BEC’s atoms instead of the interference of their wave-functions has been proposed in \cite{bravo2020phononic}.}

In spatially separated (split by a secondary base-line) CAIs, the clouds of cold atoms, used to extract local gravitational accelerations, are not exactly in the same coherent quantum states. This sets a limit on potential sensitivity of the cold atom gravity gradiometers reported to date. In BEC-based instruments, the atomic clouds are cooled down to much lower temperatures (~ 100 pico-Kelvin compared to the micro-Kelvin temperatures in standard CAIs). At such low temperatures the atomic matter forms a quantum condensate where all atoms are in the lowest possible quantum state. Two such condensate formations, typically in a spherical shape and a millimetre scale, will form perfectly matched internally entangled quantum test masses. The price to pay is the extreme complexity of dealing with such kind of delicacy as BECs. As this has been correctly stated in \cite{Szigeti2021}, any practical instruments based on BECs will need to be capable of highly precise, stable measurements in compact, low-weight and low-cost configurations, that can also operate in real-world field conditions.

Another interesting departure from either standard CAIs or BECs has been discussed in \cite{Millen:2020te,Rademacher2019QuantumSW}. Following the first demonstration of a levitated nanosphere cooled to a quantum ground state \cite{Delic:2020us}, the authors discuss a macroscopic quantum sensing “when bigger is better”. The atomic test masses are now replaced with optically controlled levitated nanoparticles trapped in optomechanical cavity. To date, there have been no reports in the public domain on using this kind of sensing for gravity gradient measurements. These developments are in their embryonic stage and seem are not capable of providing any practical instrumentation in the nearest future. 

We should highlight, there is a useful application for stationary quantum gravity gradiometers based on quantum test masses (atoms) that are the same over the age of the Universe. Providing metrological equipment and calibration stations, where local gravity gradients are measured, would allow a calibration of moving-base gravity gradiometers in absolute units and this will be a significant support for the terrestrial applications of gravity gradiometry. The in-situ gravity gradients should be determined and continuously monitored with at least 0.1 Eotvos resolution \cite{Evstifeev:2017uy}.

\section{Concluding remarks}

Finalising the above one can conclude the following: Being able to manipulate quantum objects and make them act as quantum test masses is a ground breaking research. Quantum gravity gradiometers may have promise. Right now, none of the quantum technologies that have been developed to date provide $\sim10 E/\sqrt{Hz}$ (the state-of the-art mark) sensitive gravity gradiometer at about $10-30$ cm baseline scale capable of operating in the real-world environment. For terrestrial applications, Research and Development (R and D) ceased from being profoundly active a few years back, so breakthrough ideas are needed. Quantum Gravity Gradiometers have been designed for future Space Missions. They are still having significant financial support and highly skilled academic workforce for the necessary R and D. Cold-atom and matter-wave quantum gravimeters can provide good absolute gravitational acceleration measurements under stationary conditions (on land) or on marine platforms if deployed with linearly and angularly stablised isolation systems.

\section{Addendum}

In between 2015 – 2018, three major players in the field of gravity gradiometry have left the race for creating the superconducting gravity gradiometers that could compete with the traditional Falcon and FTG generations of instruments developed by Lockheed Martin Corporation. The ARKeX, a British joint venture between Oxford Instruments (a developer of a gravity gradiometer technology based on two separated superconducting accelerometers) and ARK Geophysics (UK), went bankrupt in 2015. Its ex-executives have formed Austin Bridgeporth, a company that currently has got a license from Lockheed Martin to enhance and fly the airborne gravity gradiometers such as Air-FTG and eFTG versions of the latter. Rio Tinto terminated its long-standing financial backing of the VK-1 gravity gradiometer in 2017 that had been under development at the University of Western Australia since the 90th of the last century. Finally, Gedex Inc. went insolvent in 2018 just after reporting on the flight testing of its HD-AGG™ instrument and getting a reasonably good data. Namely in this period of time (2015-2018) there had been a peak in reporting on cold-atom and matter-wave interferometry developments over the globe and in the expectation that the shape of gravity gradiometry and its strategic terrestrial applications will be completely changed by in-coming quantum technologies and quantum sensing.

\section{Acknowledgements} 

The authors are thankful to Tom Meyer (Lockheed Martin Corporation) and Mark Dransfield, formerly a Chief Research Manager at Fugro Airborne Surveys (Australia) and later at CGG-Multiphysics (Australia), for reading this manuscript and providing valuable comments and suggestions. Funding was received from the ARC Centre of Excellence for Engineered Quantum Systems, CE170100009.

\end{document}